# DESIGNING SECURE CLUSTERING PROTOCOL WITH THE APPROACH OF REDUCING ENERGY CONSUMPTION IN WIRELESS SENSOR NETWORKS

Sanaz Sadeghi[1] and Behrouz Sadeghi[2]

[1]Department of Computer Engineering, Ferdowsi University of Mashhad, Iran

`Sanaz.sadeghi.stu@gmail.com`

[2]Department of Computer Engineering, Payam noor University of Taybad, Iran

`Sadehghi.b@pnurazavi.ac.ir`

*ABSTRACT*

*In recent years, many researchers have focused on wireless sensor networks and their applications. To obtain scalability potential in these networks most of the nodes are categorized as distinct groups named cluster and the node which is selected as cluster head or Aggregation Node offers the operation of data collection from other cluster nodes and aggregation and sending it to the rest of the network. Clustering and data aggregation increase network scalability and cause that limited resources of the network are used well. However, these mechanisms also make several breaches in the network, for example in clustered networks cluster head nodes are considered Desirable and attractive targets for attackers since reaching their information whether by physical attack and node capturing or by other attacks, the attacker can obtain the whole information of corresponding cluster. In this study secure clustering of the nodes are considered with the approach of reducing energy consumption of nodes and a protocol is presented that in addition to satisfying Advanced security needs of wireless sensor networks reduces the amount of energy consumption by the nodes. Due to network clustering there is scalability potential in such a network and According to frequent change of cluster head nodes load distribution is performed in the cluster and eventually increase the network lifetime.*

## 1. INTRODUCTION

Nowadays, wireless sensor network are increasingly becoming important in different military and civilian applications. With the recent advances in wireless communication technologies, Wireless sensor networks had a huge leap forward compared with traditional sensor networks. A wireless sensor network has a number of wireless nodes which are set on an unprotected area near or inside the targets that want to see their status as dense. These wireless nodes alternatively sense the targets' status, process the obtained data and then transfer them to a base station. All sensor nodes cooperate with each other to make a communication network in order to present reliable network services. Cooperation between sensor nodes is important in WSN due to 2 reasons:

1- Data gathered by some sensor nodes can provide a valuable inference about the environment.
2- Cooperation between sensor nodes can create compromise between Communication costs and computation energy.

Since it may be possible that data obtained from sensor node be completely coordinate with its neighbors' data, data aggregation can reduce extra information transmitted in the network. We know that energy consumed for the transfer of one bit data is equal to energy required for a number of computational operations in a sensor processor. When the base station is far applying regional data aggregation instead of direct communication has considerable advantages. Also nodes' clustering creates a topology control approach to reduce transmission overheads and applies data aggregation among a number of sensor nodes; therefore it is vital for facilitating the practical layout and performance of wireless sensor networks.

Since wireless sensor nodes are usually arranged in open and unprotected environments which don't have physical protection, they are inherently more vulnerable to malware attacks particularly in enemy environment like Battlefield traditional security mechanisms are not directly available for these networks because their Unique constraints are not considered. It is very difficult to design security mechanisms or protocols which are both efficient in the case of energy and powerful in the case of security to protect sensor network against malware attacks. In this case secure clustering means Robustness against attacks. Based on the amount of resources and knowledge of the attacker, these attacks may have different goals. For example the primary goal may be the reduction of the efficiency of gathering data sensed by sensor nodes or complete disrupting in the operation of sensor networks. This goal can be fulfilled by whether preventing algorithm ending or confusing nodes about their responsibilities in the network. Mutually, the goal of an attacker may be predicting network structure, namely predicting cluster head nodes or nodes which may be allocated to a cluster head node as a member. This may be possible by for example hearing the packages sent by nodes during the process of cluster head selection or by extracting information from parameters which are generally available. Received information can be used for attacks which target cluster heads or nodes. Of course it is clear that the attacks which target cluster head are more effective than those which attack nodes randomly. Obviously, since sensor networks often work in Hostile environments none of them can neglect the security viewpoint in designing protocol of selecting cluster head. Also, as sensor nodes are nourished by battery they have very limited energy capacity. These limitations cause many challenges in the field of developing hard ware, software, designing architectures and network protocols for sensor networks. To increase effective lifetime of sensor network, energy efficiency should be considered in all aspects of designing sensor networks. Organizing nodes in separate groups for preventing extra data transfer is a method for reducing nodes' energy consumption but on the other hand the problem of clustering is that cluster heads requiring operations such as gathering data from other nodes and their aggregation and sending them to base station have the most energy consumption so some protocols are suggested to overcome this problem in which by Intermittent change of cluster heads the amount of consumed energy is distributed in a cluster and Workload and the most energy consumption won't be limited to aggregate nodes or cluster heads.

On the other hand, based on their application wireless sensor network nodes are arranged in the environment randomly and without exact engineering and scheduling and network topology is continuously changed due to Node failures, node Damaged, adding new nodes to the network node energy drain or channel fading so we should also consider this condition that in some applications organizing nodes in Fixed and static clusters without considering network dynamism make designed protocols impractical and useless. Therefore in designing protocol for wireless sensor networks some arrangements should be considered in this field.

## 2. Related Works

Investigating the history of this subject and considering the properties that a protocol should have in order to be more secure and considering offensive models defined for different types of enemies, the protocols raised in the field of secure clustering are organized based on how much they satisfy these needs and how many mechanisms are generally used to satisfy these needs.

Protocols proposed in [2, 3, 4, 5, 6, 7, 8, 9, 10, 11, 12, 13] satisfy the properties of completion, evolution, adaptation and non manipulation of role and allocation Facing Passive attacker but they have Weaknesses facing active and aggressive attackers and in the case of these attacks just the properties of completion and evolution are met. As an example of such protocols Reliable clustering protocol suggested in [8] provides a hierarchy of clustering against cluster head failures in underwater sensor networks. This proposed algorithm is applicable not only in underwater sensor networks but also in all wireless sensor networks. This protocol selects a primary cluster head and a Backup for each cluster member to prevent re clustering in the case of failure. The network has Homogeneous sensor nodes which are located in one of the following conditions during clustering process: cluster head, cluster member or candidate for cluster head. At first each node is in the state of being candidate for cluster head. Protocol is divided in to 3 steps: Initialization, clustering and completion. In the initialization step each node finds its one hop neighbors and keeps a set of discovered neighbors which include neighbors which are still in the state of being candidate for cluster head. In this set each node creates all possible cluster heads combinations, namely if each node has n discovered neighbor it will create $2^n$ potential cluster. Among all potential clusters each candidate selects one cluster as its Eligible cluster with the average minimum cost and use a Function based on energy consumption and energy remained from cluster head members to calculate its costs. In addition to the protocols' capabilities discussed in the previous part, the protocol proposed in [14] also provides compatibility feature in facing active attacker and Allocation indiscrimination in facing both active and passive enemies. In [14] the writers proposed SecLEACH as Promotion protocol of the known clustering LEACH [15]. The main idea of LEACH is to validate the cluster head advertising messages by the help of base station and using Massage Authentication Code for all messages exchanged in the protocol. Other methods used are One-way key chains with delayed key disclosure suggested in [16] and Prior distribution of the random key suggested in [17]. Some protocols like those suggested in [18, 19, 20, 21] in addition to the capabilities of protocols of 2-2, also satisfy the feature of unpredictability of role in facing passive attacker although this set of protocols don't satisfy the features of compatibility and Allocation indiscrimination in facing active enemies.

In the protocol proposed in [19] it is supposed that all nodes are homogeneous and simultaneously connect each other under a model and time is divided in to rounds. In this protocol each message's length is limited to O(log n) and can contain Fixed number of node identifiers. The algorithm is divided in to 2 phases. In the first phase a dominant set of cluster heads are selected and in the second phase this set is expanded to a k times dominant set in which each node is covered by at least k cluster heads. The first part of the algorithm is done by repetitive reduction of active nodes. In the second part of the algorithm the number of neighboring cluster heads of all nodes are investigated. If this number c is less than k, the specified node selects kc normal neighboring nodes randomly to become its cluster heads and inform these nodes about this event. Otherwise, if sufficient number of cluster heads were selected, the algorithm is finished. The protocol proposed in [22] satisfies all criteria facing passive attacker and also facing active attacker satisfies compatibility and unpredictability. The writers suggested a cluster head selection protocol in [22] which aims at preventing that a passive attacker predicts and diagnoses the role and allocation of nodes. The general idea of this suggested algorithm is that Observable behavior of nodes is identical from the view of all nodes, this idea is attained by coding all messages of the protocol and it is necessary that all nodes send same number of messages by a random combination. Therefore, for a passive attacker no node is distinguishable from other nodes which means that the protocol doesn't leak out any information about the role and the way of nodes' allocation to cluster heads. The protocols proposed in [23, 24] are the Most advanced security protocols provided. Protocol [24] even facing an aggressive attacker satisfies compatibility conditions and provides protection about the information of the nodes' role. An example of these protocols is [23] which hides the identity of aggregated node from those who Eavesdrop from outside and the nodes which have been captured. Two data aggregation and query protocols have been also proposed which let aggregated nodes receive and aggregate the information of other nodes and send data to the main station without leaking the information out of network. Secure data aggregation protocol is formed of 4 phases: primary preparation, in this phase the required secure communication channel is created. The phase of selecting aggregating node, in this phase we should assure that no cluster is without aggregated node. Data aggregation phase which should be able to send information to aggregating node without letting other nodes understand which node is aggregation node. And query phase in which for example a mobile operator may decide to do inquiry on data.

## 3. Existing challenges

With the developments of wireless sensor networks and increasing of usage and performance of these networks in different applications, establishing security in several usage of wireless sensor networks is undeniable, for example in a network of gathering sensitive military information, if these information were accessible to all and if these information were exchanged without applying security mechanisms, in practice this kind of network is not practical.

In the existing protocols in secure clustering, some of the exclusive characteristics of these networks that are not negligible are not considered, for example, hard energy constraints, and memory limitation of sensor nodes, so defining a secure clustering protocol with considering these constraints is very necessary.

In most existing algorithms, since the cluster head elects in probabilistic manner, maybe the elected cluster heads are so near to each other so we should consider some special parameters in this decision process.

With considering the last protocols like LEACH [15] in that the nodes remaining energy was not considered in election method, maybe nodes with little remained energy becomes cluster head that causes decreasing in network lifetime.

According to probability election of the cluster heads, it is possible that the nodes in edge of the network or in regions that the density of nodes is low, selects as cluster heads so that the other nodes associated with these nodes should use more energy to send/receive data from these cluster heads.

## 4. Assumptions, limitations

A wireless sensor network is usually made of a number of multi-purpose cheap sensor nodes with low power which are arranged in the desired region. These sensor nodes are small in size but are equipped with embedded microprocessors and radio transmitters and receivers so they have not only sensor ability but also the ability of data processing and communication. They communicate in short distances by wireless media and work together for a common duty like environmental monitoring, war zone monitoring and controlling industrial process. Wireless sensor networks have some unique properties and limitation compared with traditional wireless communications networks:

*Supplying energy for sensor nodes by battery.* Sensor nodes are usually nourished by battery. Most of the times sensor nodes are arranged in harsh environments or war zones so changing or recharging of their batteries is difficult and even impossible.

*Severe limitations in energy, computations and memory.* Sensor nodes are very limited in the case of energy capacities, computations and storage.

*Configurations.* Sensor nodes are usually arranged randomly and without exact scheduling. After primary arrangement, sensor nodes should be able to configure themselves independently inside communication network.

*Unreliable sensor nodes.* Sensor nodes are usually arranged in harsh environments or war zones without supervision and are vulnerable to physical damages.

*Frequent topology change.* Network topology is frequently changed due to node failure, node damage, adding new nodes to network, node energy drain or channel fading.

*Data redundancy.* In most of sensor network applications, sensor nodes are arranged in the desired region and work together for a common purpose. Therefore data sensed by several sensor nodes surely have a level of redundancy or Solidarity.

*Numeral data.* Since in most of the applications of sensor network sensor nodes are distributed in the environment in order to monitor it and gather information about the region's status, in designing this protocol it is supposed that sensed data are numeral and sense and send quantities such as temperature, humidity and alike.

*Simulation.* Due to high cost and Unavailability of required equipments and time limitation for practical Implementation of the proposed protocol, finally the result is investigated as simulated.

The above mentioned items are unique properties and limitations which should be considered in the field of protocol design for sensor networks.

## 5. The secure clustering protocol

With considering the challenges and specific limitations of the network, the proposed secure clustering protocol consists of four phases of preparation, aggregator selection, data aggregation and data gathering.

### 5.1 The first phase-preparation

This phase is the initiation of the algorithm execution and has three steps. The target of this phase is preparation, establishing a secure channel and determining the communication network topology.

The first step of this phase is distribution of keys among the nodes. These keys are used for having secure communication among nodes and base station. In this algorithm, to establish a secure node to base station communication, each node has a common key with base station.

Also for authenticating the broadcast messages, the one way key chain is used. In one way key chain $\zeta$ is the generator that only the base station knows its value. From this generator and the production function f the chain of keys produce.

This function has the one way property, it means that we have:

$$F(k^{j-1}) = k^j \quad (1)$$

So for authenticating the base station broadcast messages, $k^j$ is sent and because the f function is saved in all nodes, with finding the output of the function with previous value of $k^j$ and comparing with present $k^j$, the nodes check the authentication of the base station.

Because with considering the nodes statically, they may be forced to send their sensed data to nodes far from them, it can cause wasting lots of energy; the second step of this phase is formation dynamic clusters. In this step, some random nodes broadcast some identifiers. The Nodes receiving this message, store the strongest signal strength of the received identifier, this identifier shows that the nodes with same identifier are closer and better selections for formation of the same cluster. In this step actually the closed nodes are identified and at last of this step all the nodes send their identifier securely to the base station with their common key with base station.

In the third step of the first phase, nodes organized in a logical ring. A ring is a path between the nodes in which every node is visited at least one time. It is possible in large rings with lots of nodes that some nodes be visited more than once that causes communication overhead in the network. However, in this protocol our idea is not restricted to the way of ring creation and in this step we just need that a logical ring being established between the closer nodes specified in previous step. So each node should just know his next hop that is his right neighbor in the ring and this neighbor must be accessible from that node.

### 5.2 The second phase- aggregator election

The main goal of this phase is that from the nodes within a ring, one or more nodes be selected as the data aggregator of the cluster and collect, aggregate and store the sensed data of the ring, while the identity of these nodes to the rest of the cluster nodes and the base station remains unknown. This phase consists of two steps. In first step, each node decides whether it wants to be an aggregator or no locally. In previous works in this field, this decision was made just in probabilistic manner, while with strong energy limitations of sensor nodes, in this protocol there are parameters that help this decision be made better.

In this step, one node which is chosen randomly starts calculating its chance of being aggregator. To do this, it sends the triple of its id, its remaining energy and a counter that at first is set to one, to its right neighbor in the ring. Each node with receiving this message adds its remaining energy to the one in the message and increases the counter one unit and forwards it to its right neighbor. This process continues in the ring till the message arrives to the starter node (the initiator id is sent over the message). This node calculates the average remaining energy of the ring by dividing the sum of the remained energy to the counter and forwards this value one round more in the ring. Every node calculates its chance of being aggregator by this value proportionally, so to increase throughput and prolonged network lifetime, this election is done not probability but with considering the specifications of the node. The equations below show these processes.

$$E_{avg} = (e/counter) \quad (2)$$

$$Pr \sim (e - E_{avg}) \quad (3)$$

In the second step of the second phase, in order to prevent of not having aggregator in the cluster, the nodes must run a specific process that make sure that in this ring there is selected at least one aggregator node and also the identity of this(these) aggregator node(s) are still unknown. This step can be done by the protocol presented in [25].

### 5.3 The third phase- data aggregation

In this phase the process of how data is forwarded in the ring without knowing the identity of the aggregators is specified. This procedure uses directly the ring topology which was formed in preparation phase. The starter node starts the aggregation by sending its sensed data as a token to its right neighbor. With receiving the token each node, adds its data to the token and forward it to its neighbor. When the starter receives the token again, it forwards this value again in the ring and so the aggregator(s) can save the aggregated data.

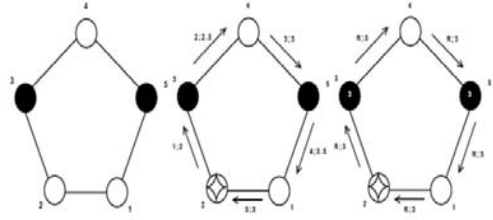

Figure1. Data aggregation

### 5.4 The fourth phase- data collection

The final goal of a sensor network is forwarding sensed data to the base station. This sending may be on demand of the base station that is called query-driven. The purpose of this protocol is that the requested data of the base station being provided while aggregator node's identity still remains unknown. In the solution presented in this issue, it is need that the base station places in the frequency range of one of the nodes inside the ring. To establish a secure connection and to authenticate the base station to each node, a one-way key chain that was described in the first phase is used. After base station selects one node randomly, it sends its query to the selected node. Once, for example, node A become sure of the identity of the base station forwards the query into the ring, each node receives the query if it is a non aggregator node just encrypt a nonsense data and forward it and if the node is aggregator it also adds the aggregated data to it. This process can be summarized as follows:

$$R_i = \begin{cases} h(Q|k_i), & \text{for non aggregators} \\ h(Q|k_i) + M, & \text{for aggregators} \end{cases}$$

Figure2. Data collection function

With receiving the message from node A, the base station can easily reproduce the hash values, because it shares key with all the nodes in the ring. So with subtracting the hash values from the final R, cM is got where c is the number of aggregator nodes within the cluster and M is the measured data. Given that the base station doesn't know the number of aggregator nodes in the ring, with selecting the appropriate value of c, the original c can be found. With respect to the range of measured values, it is easily proved [23] that in this area there is only one value for c that we have:

$$c = \hat{c}$$

## 6. EVALUATION RESULTS

At the end to compare the proposed protocol in this paper with other protocols presented in this field, we use evaluation metrics that were introduced in [1].

Relating to security, and in order to find how much secure this protocol is, the below parameters are used.

### 6.1 Termination

The termination property requires the protocol to terminate in a limited time, so in this way to meet this property dead-locks or infinite waiting times are not allowed. In this protocol since the selection process lasts in finite time, it meets this property.

## 6.2 Completeness

The completeness property can be interpreted only if the termination property is met, and requires every terminated node to have its role.

## 6.3 Consistency

According to the consistency property, if the node's role is that of a cluster member, then the node should be associated with a cluster head that further believes itself as cluster head. As a counterexample, if a cluster member node thinks that it is associated with node B, and node B does not believe itself to be cluster head, then the protocol is not consistent.

## 6.4 Non-manipulability

Two different kinds of non-manipulability property can be defined, role non-manipulability and association non-manipulability. These properties state that the adversary should not be able to alter the role or the association of the sensor nodes during the election process. This property is important, because if an adversary could alter its role or association, then he could, for example, force a node controlled by him to become cluster head all the time. Hence, he could take over the control of a significant part of the sensor network and the whole cluster of that cluster head. And also, the enemy could force it's under controlled nodes to always become a cluster member, thus it can help the enemy for saving energy of its nodes.

## 6.5 Unpredictability

Unpredictability makes the adversary incapable of precisely determining the identity of the cluster head before the election process; this goal is to prevent the adversary from predicting the upcoming cluster head, since that node may become the primary target of attacks.

## 6.6 Unidentifiability

Unidentifiability is very similar to unpredictability with this difference that it allows the adversary to observe the election process and nevertheless the enemy would not be able in detecting the roles of the nodes and also the association of the nodes that means which node is belong to which cluster.

With introducing these parameters and with analyzing our presented protocol, we can claim that this protocol meets all the security requirements and have met the six properties defined so far with respect to all kinds of adversaries like passive adversary, active adversary and compromised enemy.

## 7. CONCLUSIONS

The principle of the proposed protocol is to establish a secure network and provides a secure protocol in wireless sensor networks, and this protocol works unlike previous approaches that only consider security as an important issue, and they did not consider the severe limitations of sensor nodes. These regardless of the characteristics of sensor nodes, would result that these secure protocols don't have good performance in practice, so with proposing this protocol, high security of the network in special and sensitive applications is completely provided and also the energy consumption of the nodes is considered specially.

In the previous protocols it was assumed that the clusters that are going to integrate their data securely are static, while this assumption imposes sever limitations to the network in practice and is inconsistent with lots of applications of these networks that the nodes are distributed in the environment randomly and unplanned. Because in previous approaches the nodes that were known as the neighbors statically, maybe place very far from each other after deployment, but they are forced to have communication because they are their right neighbor. So in this protocol in the second step of the first phase, dynamic clustering is fully presented to solve this problem.

In the second phase of the proposed protocol which is aggregator election, the idea is that the nodes within a cluster make decision that whether they want to be aggregator or not with considering consciously, not just by probability. They consider their remaining energy in making this decision and in this case this protocol besides to high security, consider the limitations of the network and so help prolonging the network life. And also with studying many of the techniques in this regard, in presenting this protocol the top features for securing a cluster formation in face of active, passive and even compromised adversary is considered and is tried that in some steps of the protocol, some of these ideas being used. So this protocol is a combination of the best ideas with new our new approaches to fix their defects. Thus in this protocol besides the lower energy consumption and longer network lifetime, there are no information leakage to the out of the network and all the messages are sent confidentially and all the senders of the messages are also authenticated.

## ACKNOWLEDGEMENTS

I would like to express my deep gratitude to Professor Mohammad Hossein Yaghmaee Moghaddam and Dr. Hossein Deldari, my research supervisors, for their patient guidance, enthusiastic encouragement and useful critiques of this research work.

I would also like to extend my thanks to the technicians of the IPPBX laboratory of the Computer department of Ferdowsi University of Mashhad for their help in offering me the resources in doing the work.

Finally, I wish to thank my parents and my dear husband for their support and encouragement throughout my study.

**Sanaz Sadeghi** was born on June 1988 in Mashhad, Iran. She received the BS degree in computer engineering from the computer department, Ferdowsi University of Mashhad, Mashhad, Iran, in 2010. She is a MS student in computer engineering at computer department, Ferdowsi University of Mashhad. Her research interests are in computer networks, wireless sensor networks and network security.

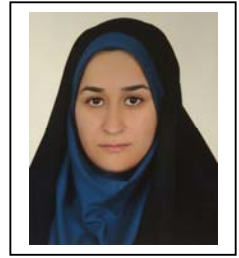

**Behrouz Sadeghi** was born on June 1982 in Mashhad, Iran. He has the MS degree in computer software engineering. He is now a teacher in department of computer and Information Technology in faculty of computer Software, Payam Noor University of Taybad, Iran. His research interests are in distributed computing, grid resource management and wireless sensor networks.

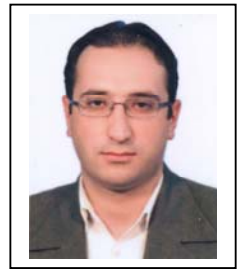